\documentclass[letterpaper, 10 pt, conference]{ieeeconf}  

\IEEEoverridecommandlockouts                              

\usepackage{amsmath,amsfonts}
\usepackage{algorithmic}
\usepackage{algorithm}
\usepackage{array}
\usepackage[caption=false,font=normalsize,labelfont=sf,textfont=sf]{subfig}
\usepackage{textcomp}
\usepackage{stfloats}
\usepackage{url}
\usepackage{verbatim}
\usepackage{comment}
\usepackage{graphicx}
\usepackage{cite}
\usepackage{siunitx}
\usepackage{balance}
\usepackage{color}
\usepackage{booktabs}
\usepackage{multirow}
\usepackage[normalem]{ulem}

\newcommand{\revision}[1]{{\color{black}{#1}}}

\begin{document}

\onecolumn
\textcopyright~2024 IEEE.  Personal use of this material is permitted.  Permission from IEEE must be obtained for all other uses, in any current or future media, including reprinting/republishing this material for advertising or promotional purposes, creating new collective works, for resale or redistribution to servers or lists, or reuse of any copyrighted component of this work in other works.
\newpage
\twocolumn
\title{}

\title{\LARGE \bf
Reliability of Smartphone-Based Vibration Threshold Measurements

\thanks{$^{1}$R.\ A.\ G.\ Adenekan, K.\ T.\ Yoshida, and A.\ M.\ Okamura are with the Mechanical Engineering Department and A.\ Gonzalez Reyes is with the Computer Science Department at Stanford University, \ Stanford, CA 94305. (email: \{adenekan; kyle3; aokamura\}@stanford.edu; alegre@cs.stanford.edu).}%
\thanks{$^{2}$A.\ Benyoucef is with the Mechanical Engineering Department at the University of Colorado Denver, Denver, CO 14850. (email: anis.benyoucef@ucdenver.edu).}%
\revision{\thanks{$^{3}$A.\ E.\ Adenekan, Houston, TX 77083. (email: aadenekan@yahoo.com).}}%
\thanks{$^{4}$C.\ M.\ Nunez is with the Sibley School of Mechanical and Aerospace Engineering at Cornell University, Ithaca, NY 14853. (email: cmn97@cornell.edu).}%
\thanks{*This work was supported in part by pilot grants from the Precision Health and Integrated Diagnostics Center at Stanford, \revision{the Stanford Center for Digital Health, and the Stanford Diabetes Research Center (pilot grant \#P30DK116074). The work was also supported by }the National Science Foundation Graduate Research Fellowship Program, the Stanford Graduate Fellowship Program, and the Stanford Summer Undergraduate Research Fellowship Program (SURF).}
\author{Rachel A.\ G.\ Adenekan$^{1}$,~\IEEEmembership{Student Member,~IEEE,}, Kyle T.\ Yoshida$^{1}$,~\IEEEmembership{Member,~IEEE,},\\ Anis Benyoucef$^{2}$, Alejandrina Gonzalez Reyes$^{1}$, Adeyinka E. Adenekan$^{3}$,\\ Allison M.\ Okamura$^{1}$,~\IEEEmembership{Fellow,~IEEE,}  and Cara M.\ Nunez$^{4}$,~\IEEEmembership{Member,~IEEE} }

}

\maketitle
\thispagestyle{empty}
\pagestyle{empty}

\begin{abstract}
Smartphone-based measurement platforms can collect data on human sensory function in an accessible manner. We developed a smartphone app that measures vibration perception thresholds by commanding vibrations with varying amplitudes and recording user responses via (1) a staircase method that adjusts a variable stimulus, and (2) a decay method that measures the time a user feels a decaying stimulus. We conducted two studies with healthy adults to assess the reliability and usability of the app when the smartphone was applied to the hand and foot. \revision{
The staircase mode had good reliability for repeated measurements, both with and without the support of an in-person experimenter. The app has the potential to be used at home in unguided scenarios.}

\end{abstract}

\section{Introduction}

\begin{figure}
\centering
\includegraphics[width=1.0\linewidth]{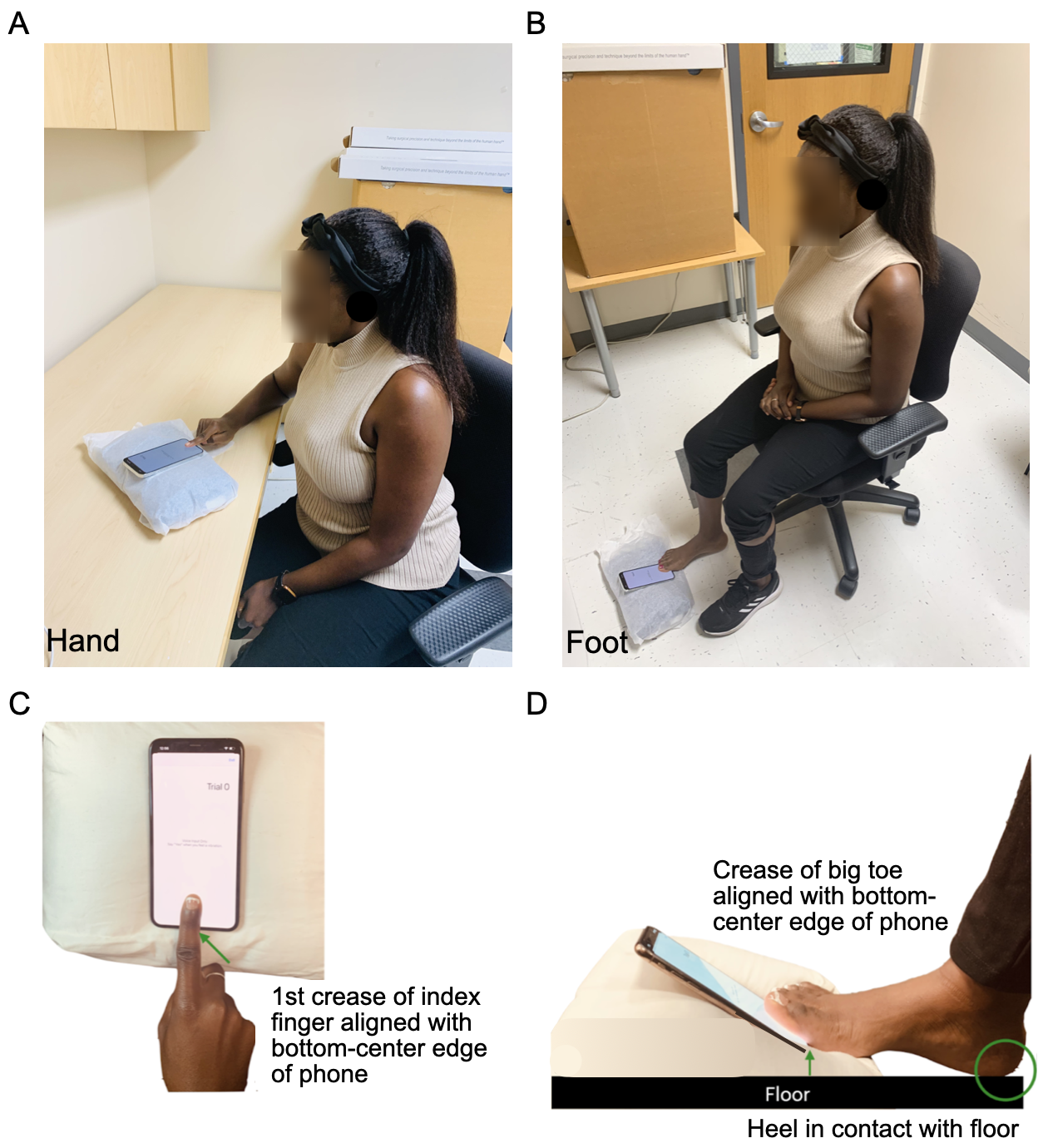}
\caption{Smartphone setup pertaining to the studies described in Section~\ref{sec:repeated} and Section~\ref{sec:guidance} for both the hand (A and C) and foot (B and D). The phone is placed on a pillow, and users wear noise cancelling headphones playing white noise. Detailed placement of the hand (C) and foot (D) placements are also displayed.}
\label{setup}
\end{figure}





\revision{Vibrotactile perception enables humans to perform sensorimotor tasks important for activities of daily living, such as object manipulation~\cite{johnson1999vibration} and walking~\cite{era1996postural}.
Vibrotactile perception diminishes with age, and peripheral neuropathy, caused by diabetes, autoimmune disorders, chemotherapy, or neurological impairments, can also result in substantial alterations~\cite{halonen1986aging, hanewinckel2016peripheral}. Diminished vibrotactile perception poses a significant risk to individuals, increasing the likelihood of infections, ulcers, and falls~\cite{o1993incidence}. Thus, it is important to measure and track vibrotactile perception, especially for individuals with or \revision{at risk of developing} peripheral neuropathy.}

A 128~Hz clinical tuning fork (CTF) exam is the standard method used to measure vibrotactile perception thresholds (VPTs)~\cite{pop2017diabetic}. To perform the exam, a clinician strikes the tines of the tuning fork, places the end of the fork on the patient, and \revision{either} counts the seconds until the patient verbally states that they can no longer feel a vibration \revision{or, more commonly, simply records whether or not the patient feels any vibrations}. This exam often results in inconsistently applied vibrations caused by differences in how the clinician strikes the fork from one use to the next~\cite{rossing1992acoustics}. Therefore, this method does not provide accurate longitudinal measurements. A measure used to more precisely measure VPTs is a 64 Hz Rydel-Seiffer tuning fork (RSTF) exam~\cite{panosyan2016rydel}. \revision{To perform the exam, the clinician presses the tines of the fork together, releases them, and places the end of the fork on the individual. When the fork is vibrating, markings on a tine create the optical illusion of a triangle, the length of which indicates the amplitude of the vibration. The experimenter reads the length of the illusory triangle on a scale between 0 and 8 at the time the patient states they can no longer feel the vibration. While this semi-quantitative approach yields more insight than the CTF,} the RSTF has not been widely adopted in clinical practice because \revision{clinicians state that} its benefit often does not justify its cost. It has mostly been used in neurology research~\cite{alanazy2018conventional}. A precise, high-resolution, and accessible measurement tool is required to effectively monitor changes in vibrotactile perception over time scales that are conducive to guiding treatment and management plans.


\revision{
Due to the ubiquity of smartphones and recent advancements in smartphone haptic hardware and controls, researchers have \revision{
expressed interest in} using smartphones to conduct mobile haptics experiments}~\cite{blum2019getting,yoshida2023cognitive,yoshida2023responseTimes}, and specifically \revision{
to measure} VPTs~\cite{adenekan2022feasibility,torres2023skin,LindsaySubmitted}. In our recent work, we present a smartphone app platform that utilizes psychophysics, specifically the staircase method, to measure VPTs~\cite{AdenekanAccepted2024}. The results from this prior work indicate that we can measure VPTs with our smartphone app platform which are correlated to current clinical methods. However, to realize the full potential of this platform to track tactile perception and possibly even measure disease progression and regression outside of a clinic, it is imperative that the measurements from our tool are also consistent, reliable, and able to be conducted by individuals without guidance from an experimenter or clinician. In this work, we expand upon our smartphone app platform and explore this topic by conducting two human subject user studies. In Section~\ref{sec:repeated}, we present our smartphone app which can calculate VPTs using the previously developed smartphone staircase method and a new smartphone decay method \revision{
that mimics} the applied vibrations from a CTF or RSTF. We also present the first human subject study we conducted in which we evaluated the reliability of our methods compared to current clinical methods by collecting repeated measurements\revision{
}. Next, in Section~\ref{sec:guidance}, we discuss the second human subject user study we conducted which explores if our platform can collect similar VPT measurements when participants do not receive guidance from an experimenter. Last, we discuss the conclusions from our work and future directions in Section~\ref{sec:conclusion}.




\section{Study 1: Reliability of Repeated Measurements}
\label{sec:repeated}
The purpose of this study was to quantify how reliably our platform measures VPTs. To that end, we measured how VPTs change between three repeated trials measured via our platform (smartphone staircase and smartphone decay) and via clinical benchmarks (RSTF and CTF) at two body locations (the hand and the foot) in 11 participants and compared VPTs of repeated measurements for each condition. We describe the smartphone and tuning fork measurement methods, user study protocol, and statistical analysis, followed by results and discussion.

\subsection{Methods}
\subsubsection{Smartphone Measurement Methods}
\label{subsec:phone meas methods}

\begin{figure*}[b]
\centering
\includegraphics[width=1.0\linewidth]{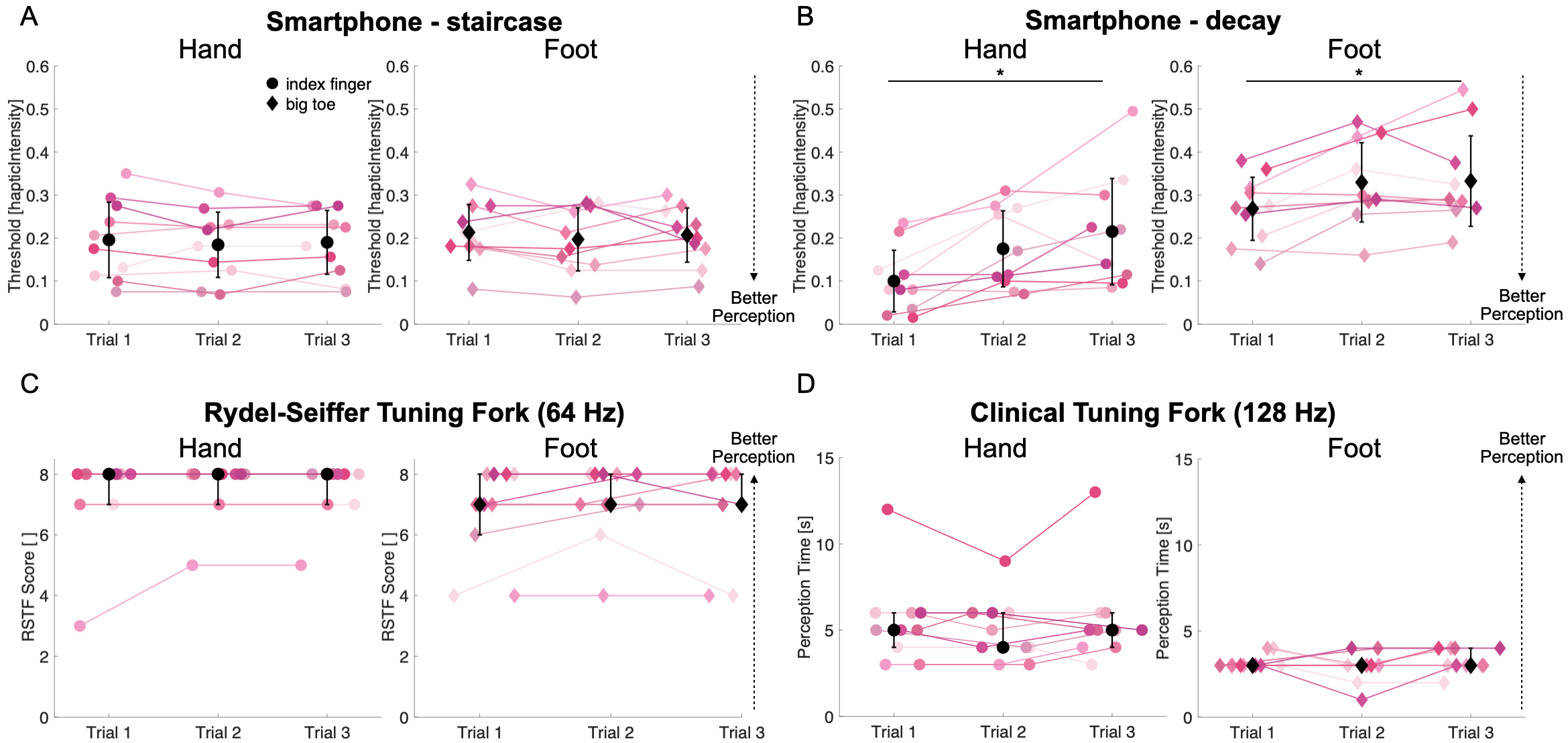}
\caption{Repeated threshold measurements for each tool on Day 1. Individual participants \revision{($n=10$)} are shown in shades of pink, and group level data is shown in black. For the smartphone methods (A and B), mean and standard deviation are shown. For the tuning fork methods (C and D), median and quantiles (25th and 75th) are shown. Unlike the other tools, the smartphone decay method (B) had significant variability between each of the trials on a single day for both the hand and foot positions. Only Day 1 data is plotted in this figure given space limitations and lack of major differences between Days 1 and 2. Statistical results for both Days 1 and 2 are provided in Table ~\ref{Table: stats}. \revision{Although we tested 11 participants, we only present data from 10 participants. One participant's results were not used in the analyses as explained in Section~\ref{subsec:study 1 stat methods} }} 
\label{test-retest}
\end{figure*}

We conducted all smartphone VPT tests on an Apple iPhone XS Max, in which the amplitude of the vibration is modified using both the staircase method described in prior work~\cite{AdenekanAccepted2024} and a newly designed decay method. The decay method is based on the exponentially decaying vibration pattern created from a clinical tuning fork\revision{
}. The frequency of the phone vibrations was held constant at 230~Hz (\textit{hapticSharpness} = 1), while the variable controlling the amplitude of the phone's vibration, \textit{hapticIntensity}, was autonomously and continuously linearly lowered from an initial value of \textit{hapticIntensity} = 1 at time = 0 to \textit{hapticIntensity} = 0 at time = 20 seconds. This continuous decay happened over 20 seconds (which is similar to the amount of time it takes a 128 Hz clinical tuning fork's vibrations to decay after being struck. Both \textit{hapticIntensity} and \textit{hapticSharpness} are variables within Apple's Core Haptics Framework that can be set to any number between 0 and 1. We chose \textit{hapticSharpness} = 1 (230 Hz) and \textit{hapticIntensity} = 1 (0.65g) because those are the maximum frequency and vibration acceleration amplitude output by the smartphone we used (iPhone XS Max)~\cite{AdenekanAccepted2024}. 230 Hz is also within the stimulation range of the Pacinian corpuscles which play a key role in vibrotactile perception at high frequencies~\cite{verrillo1992vibration}. As described in~\cite{AdenekanAccepted2024}, linear decays in \textit{hapticIntensity} result in exponential decays in vibration amplitude (in units g) so by linearly decaying the \textit{hapticIntensity}, we are exponentially decaying the vibration amplitude.

In the decay exam, users were instructed to say ``start'' when they were ready to begin the exam. The app uses Apple’s Speech Framework, which can recognize spoken word from recorded or live audio, to identify this command and trigger the start of the vibrations. Users were instructed to say ``stop'' when they could no longer feel the vibrations. Users who did not immediately feel a vibration after saying ``start'' were instructed to say ``nothing''. Upon recognizing the ``stop'' command, the app recorded the current value of the \textit{hapticIntensity} and ended the trial. Upon recognizing the ``nothing'' response, the app recorded that the user could not feel the vibrations and ended the trial. After completion of all the trials, a \texttt{.csv} file containing the trial data was automatically saved via Google Firebase. Trials were repeated three times in a row at both body locations of interest (pad of the index finger and pad of the big toe on the dominant hand/foot). The smartphone VPT ranges from \textit{hapticIntensitity} of 0.05 (smallest vibration detected) to 1 (largest possible vibration output via Apple's framework). As such, lower scores indicate better perception. If the largest vibrations output by the phone (vibrations generated by setting \textit{hapticIntensity} to 1) were not felt, the threshold for that trial is recorded as \texttt{NaN} because their threshold was outside the range of the smartphone-generated vibration stimuli.

Figure~\ref{setup} shows the setup for both the hand and foot positions. In both cases, the phone was placed on a pillow because placing the phone directly on the table or floor results in dampened vibrations and thus reduces the range of the tool. As shown in Fig.~\ref{setup}C, the first crease (distal interphalangeal joint) of the index finger was aligned with the bottom-center edge of the phone and the rest of the fingers were curled into a fist and rested on the table. Similarly, the crease (interphalangeal joint) of the big toe was aligned with the bottom-center edge of the phone and the heel was in contact with the floor (Fig.~\ref{setup}D). Participants wore noise-cancelling headphones playing white noise to enforce reliance on feeling instead of hearing the vibrations. 

\subsubsection{Tuning Fork Measurement Methods}
We conducted tuning fork measurements using two techniques: (1) the 128 Hz CTF exam and (2) the semi-quantitative 64 Hz RSTF exam. We tested the same two body parts as in the smartphone setup as shown in Fig.~\ref{setup}. The physical setup for both tuning forks was the same as that for the tuning fork exam described in ~\cite{AdenekanAccepted2024}.

\revision{
Following the methods described in \cite{AdenekanAccepted2024}, the experimenter conducted the 128 Hz CTF exam by striking the tines of the tuning fork (Cynamed), placing the end of the fork on the individual, and counting the seconds until the individual verbally states that they can no longer feel a vibration. A time of 0 seconds indicates that a person cannot feel the tuning fork's vibrations. As such, longer times indicate better perception.}

For the RSTF exam, the experimenter pressed the tines of a 64 Hz RSTF (US Neurologicals) together, released them, and then placed the base of the tuning fork on the body of the participant. Participants were instructed to say ``stop'' when they no longer felt any vibrations. \revision{Each participant's RSTF score was recorded as the number between 0 and 8 that was closest to the tip of the illusory triangle generated by the vibration and markings on the tines, at the time the participant said ``stop''. This follows the methods used in clinical research settings; figures that depict an RSTF in use are shown in~\cite{panosyan2016rydel}}. The RSTF score ranges from 0 (largest vibration undetected) to 8 (smallest vibration detected). As such, higher scores indicate better perception. Prior to collection, participants were touched with a vibrating tuning fork so they could get a sense of what it felt like and so that they understood that the sensation was not painful. Participants wore a blindfold so that they could not see when the tuning fork made contact with their body, but did not wear headphones as the sound of the tuning fork vibrations are not easily discernible. At each body location, the RSTF score was measured three times in a row before moving to the next body part.

\subsubsection{User Study Participants}
Eleven adult participants with no known history of diabetes or other disorders linked to peripheral neuropathy completed this study. Participant demographics were as follows: 6 reported female and 5 reported male for their sex assigned at birth; aged 21 to 27 (mean: 24.7, std: 1.9); all right-handed; 11 right- and 1 left-footed; 4 were Black/of African descent, 3 were Asian, 4 were White, 1 was Hispanic or Latino, and 1 preferred not to answer (some participants listed multiple identities). This study was approved by the Stanford University Institutional Review Board under Protocol 22514 and written consent was provided by all participants. Prior to completing the study, participants completed a pre-survey that inquired about demographic information as well as hobbies or injuries that may impact touch sensitivity at the hands or feet.

\subsubsection{User Study Protocol}
Participants completed a two-day protocol with a one-hour session each day. The same time block was used for each day. All four exams (smartphone staircase, smartphone decay, RSTF, and CTF) were conducted on both days. The ordering of the exams within each session was randomized. For each given modality, the ordering of the body parts tested was also randomized. All perception measurements were collected on the participant’s dominant side body parts.

\subsubsection{Statistical Methods}
\label{subsec:study 1 stat methods}
\revision{Smartphone (staircase and decay) and tuning fork (CTF and RSTF) thresholds for one participant were removed from further analysis due to a file-save error during the staircase portion of the experiment. This resulted in ten participants with usable data.} We aimed to assess the precision of our smartphone methods by understanding how repeatable the measurements are within a single day by collecting 3 repeated measurements of each method. We repeated the protocol and corresponding statistics on a second day and present statistics from both days separately in Section~\ref{sec:resultsRepeat}.

One-way repeated measures ANOVAs were used to analyze whether the VPT for the smartphone methods is reproducible between each consecutive trials. Friedman's Tests (the non-parametric equivalent of the ANOVA) were used to compare the effect of each trial on VPT for the tuning fork methods, due to lack of normalcy and their more discrete characteristics. The repeated measures ANOVA and Friedman's Tests were conducted in R (Version 4.2.1) using R Studio (Version 2022.07.2).

The quantile calculations for the tuning fork data displayed in Fig.~\ref{test-retest} were performed
using R allowing for quantiles to be calculated for discrete and non-normally distributed data that are unsuitable for linear interpolation. We used R’s type 1 quantile calculation algorithm.

Intraclass correlation coefficients (ICC) with two-way mixed effects, absolute agreement, single rater models are used to compare the precision (how similar each of the three trials within a day are to each other) of the two smartphone methods to each other and understand how this relates to the currently accepted reliability of the tuning forks. The ICC calculations were performed
using IBM SPSS Statistics (Version 29.0.0.0).

\subsection{Results}
\label{sec:resultsRepeat}

\begin{table*}[]
\centering
\caption{Reliability of Repeated Measurements Statistical Results}
\label{Table: stats}
\begin{tabular}{ll|llll|llll|}
\cline{3-10}
 &
   &
  \multicolumn{4}{l|}{Day 1} &
  \multicolumn{4}{l|}{Day 2} \\ \hline
\multicolumn{1}{|l|}{Location} &
  \text{Method} &
  \multicolumn{1}{l|}{P-value} &
  \multicolumn{1}{l|}{$\eta_{\text{p}}^{2}$} &
  \multicolumn{1}{l|}{W} &
  \begin{tabular}[c]{@{}l@{}}ICC (Confidence Interval)\end{tabular} &
  \multicolumn{1}{l|}{P-value} &
  \multicolumn{1}{l|}{$\eta_{\text{p}}^{2}$} &
  \multicolumn{1}{l|}{W} &
  ICC (Confidence Interval) \\ \hline
\multicolumn{1}{|l|}{\multirow{4}{*}{Hand}} &
  \text{Staircase} &
  \multicolumn{1}{l|}{0.574} &
  \multicolumn{1}{l|}{0.06} &
  \multicolumn{1}{l|}{\textbf{-}} &
  \textbf{0.924 (0.802, 0.979)} &
  \multicolumn{1}{l|}{0.335} &
  \multicolumn{1}{l|}{\text{0.11}} &
  \multicolumn{1}{l|}{-} &
  \textbf{0.951 (0.870, 0.986)} \\ \cline{2-10} 
\multicolumn{1}{|l|}{} &
  Decay &
  \multicolumn{1}{l|}{\textbf{$<$0.001}} &
  \multicolumn{1}{l|}{\textbf{0.56}} &
  \multicolumn{1}{l|}{\textbf{-}} &
  \text{0.555 (0.121, 0.851)} &
  \multicolumn{1}{l|}{\textbf{$<$0.001}} &
  \multicolumn{1}{l|}{\textbf{0.54}} &
  \multicolumn{1}{l|}{\textbf{-}} &
  0.775 (0.363, 0.938) \\ \cline{2-10} 
\multicolumn{1}{|l|}{} &
  CTF &
  \multicolumn{1}{l|}{0.368} &
  \multicolumn{1}{l|}{-} &
  \multicolumn{1}{l|}{0.1} &
  \text{0.880 (0.704, 0.965)} &
  \multicolumn{1}{l|}{\textbf{0.048}} &
  \multicolumn{1}{l|}{-} &
  \multicolumn{1}{l|}{\textbf{0.3}} &
  0.773 (0.456, 0.932) \\ \cline{2-10} 
\multicolumn{1}{|l|}{} &
  \text{RSTF} &
  \multicolumn{1}{l|}{0.368} &
  \multicolumn{1}{l|}{-} &
  \multicolumn{1}{l|}{\text{0.1}} &
  \textbf{0.908 (0.765, 0.974)} &
  \multicolumn{1}{l|}{0.264} &
  \multicolumn{1}{l|}{\textbf{-}} &
  \multicolumn{1}{l|}{0.1} &
  \textbf{0.905 (0.759, 0.973)} \\ \hline
\multicolumn{1}{|l|}{\multirow{4}{*}{Foot}} &
  \text{Staircase} &
  \multicolumn{1}{l|}{0.540} &
  \multicolumn{1}{l|}{0.07} &
  \multicolumn{1}{l|}{-} &
  0.797 (0.535, 0.939) &
  \multicolumn{1}{l|}{0.115} &
  \multicolumn{1}{l|}{0.25} &
  \multicolumn{1}{l|}{-} &
  0.733 (0.424, 0.917) \\ \cline{2-10} 
\multicolumn{1}{|l|}{} &
  Decay &
  \multicolumn{1}{l|}{\textbf{0.006}} &
  \multicolumn{1}{l|}{\textbf{0.43}} &
  \multicolumn{1}{l|}{\textbf{-}} &
  0.703 (0.323, 0.909) &
  \multicolumn{1}{l|}{0.121} &
  \multicolumn{1}{l|}{0.21} &
  \multicolumn{1}{l|}{-} &
  0.785 (0.514, 0.934) \\ \cline{2-10} 
\multicolumn{1}{|l|}{} &
  CTF &
  \multicolumn{1}{l|}{0.368} &
  \multicolumn{1}{l|}{-} &
  \multicolumn{1}{l|}{0.1} &
  0.302 (-0.060, 0.710) &
  \multicolumn{1}{l|}{0.196} &
  \multicolumn{1}{l|}{-} &
  \multicolumn{1}{l|}{0.2} &
  0.759 (0.474, 0.925) \\ \cline{2-10} 
\multicolumn{1}{|l|}{} &
  RSTF &
  \multicolumn{1}{l|}{0.247} &
  \multicolumn{1}{l|}{-} &
  \multicolumn{1}{l|}{0.1} &
  0.881 (0.704, 0.965) &
  \multicolumn{1}{l|}{0.819} &
  \multicolumn{1}{l|}{\textbf{-}} &
  \multicolumn{1}{l|}{0.0} &
  \textbf{0.912 (0.769, 0.975)} \\ \hline
\end{tabular}
\end{table*}


Fig.~\ref{test-retest} shows VPTs measured for three trials using each tool (Smartphone-staircase, Smartphone-decay, RSTF, and CTF).
Table~\ref{Table: stats} shows the results from the described statistical analyses examining changes in VPT between trials and calculating ICCs. $\eta_{\text{p}}^{2}$ shows the partial effect sizes from the ANOVAs, while Kendall's W shows the effect sizes from the Friedman's Tests. The ICCs and their corresponding confidence intervals show how similar the VPT is between trials.

\subsubsection{Between-Trial Comparisons}
ANOVAs and Friedman's Tests were used to determine if there was a significant impact of trial on the measured VPT. Significant differences between trials would indicate that the VPT fluctuates between different measurements.

\emph{Hand}: The VPT was not statistically significantly different between trials for the smartphone staircase nor the RSTF method for either Day 1 or 2. The VPT was not statistically significantly different between trials for the CTF method on Day 1, but was different on Day 2 (${\chi}^2$$(2) = 6.07$, $p = 0.048$, $W = 0.303$). However, post hoc pairwise comparisons using Wilcoxon signed-rank tests revealed no statistically significant differences between any groups. The VPT was statistically significantly different between trials on both Day 1 ($F(2,18) = 11.64$, $p < .001$, $\eta_{\text{p}}^{2}$$ = 0.56$) and Day 2 ($F(2,18) = 10.54$, $p < .001$, $\eta_{\text{p}}^{2}$$ = 0.54$) for the smartphone decay method. Post hoc pairwise comparisons for Day 1 showed statistically significant differences between trial 1 and trial 2 ($p = 0.012$) and between trial 1 and trial 3 ($p = 0.004$). Post hoc pairwise comparisons for Day 2 showed statistically significant differences between trial 1 and trial 3 ($p = 0.005$).

\emph{Foot}: The VPT was not statistically significantly different between trials for the smartphone staircase, CTF, nor RSTF methods for neither Day 1 nor 2. The VPT was statistically significantly different between trials for the smartphone decay method on Day 1 ($F(2,18) = 6.80$, $p = .006$, $\eta_{\text{p}}^{2}$$ = 0.43$), but was not on Day 2. Post hoc pairwise comparisons showed statistically significant differences between trial 1 and trial 2 on Day 1 ($p = 0.011$).

\subsubsection{Intraclass Correlation Coefficients}
ICCs are statistical measures that are commonly used to assess the reliability of clustered data. ICC values can be evaluated as follows: $>0.9$ indicates excellent reliability, $0.75-0.9$ indicates good reliability, $0.5-0.75$ indicates moderate reliability, and values $<0.5$ indicate poor reliability~\cite{koo2016guideline}. 
ICC values and their confidence intervals for each tool and body location on both days are present in Table~\ref{Table: stats}. Below we highlight some key results.

\emph{Hand}: The smartphone staircase and RSTF methods had excellent reliability at the hand on both days, with ICC values consistently above $0.9$. The decay method showed the lowest ICC value on Day 1 ($0.555$), though this was higher on Day 2 ($0.755$). 

\emph{Foot}: At the foot, the RSTF method had the highest ICC on both Days 1 and 2. The CTF method showed the lowest ICC overall at $0.302$ on Day 1. Notably, the smartphone staircase method still showed good reliability at the foot.

\subsection{Discussion}
From this user study, we aimed to evaluate the reliability of VPTs measured from a smartphone device and to understand how the reliability of smartphone VPTs compares to those typically used in clinical settings. Our results indicate that although CTFs are the current standard in clinical practice, their reliability is questionable \revision{due to variability in striking, and thus in resulting vibration amplitudes experienced by patients} (which has been alluded to in prior work~\cite{rossing1992acoustics}). The CTF ICC values were in the good reliability range for the hand on both days, but only good reliability on the foot on Day \revision{
2}; the ICC value for the CTF on Day \revision{
1} was the lowest across all conditions at $0.302$ indicating poor reliability. \revision{
The} smartphone decay method, which was designed based on the vibrations provided by CTF \revision{and RSTF}, did not produce the most reliable measurements and significantly varied across trials within a given session for the hand on both days and for the foot on Day 1 (although not Day 2). The ICC values for the smartphone decay method typically fell in the moderate range, with the highest ICC value of $0.785$ occurring at the foot on Day 2 and the lowest ICC of $0.555$ occurring at the hand on Day 1.

In contrast, VPT measurements using the RSTF and smartphone staircase methods were never significantly different within a session and resulted in similar ICC values mainly in the good or excellent reliability range. The measured ICC values for the RSTF were in range with measurements obtained in prior work~\cite{marcuzzi2019vibration}. The smartphone staircase method resulted in the highest ICC values on the hand ($0.924$ for Day 1 and $0.951$ for Day 2) all in the excellent reliability range indicating that it was the best-performing VPT measurement method. Although the ICC values on the foot for the smartphone staircase method are a little bit lower ($0.797$ for Day 1 and $0.733$ for Day 2), they still indicate that the smartphone staircase method can serve as an effective and reliable tool for measuring VPTs. One\revision{
} reason for the better performance of the staircase exam compared to the decay exam is that the recorded threshold is more likely to be affected by speech processing delays during the decay exam (the app relies on the user saying ``stop'' to trigger the recording of vibration amplitude detected). In contrast, the staircase app provides users with a dedicated response window of $2.5$ seconds before providing the next vibration stimulus. Furthermore, thresholds measured via increasing stimuli are known to perform better than those measured via decaying stimuli, and the staircase exam essentially averages increasing and decreasing stimuli~\cite{goldberg1979standardised}. 

To conclude, our results indicate that measuring VPTs using our smartphone staircase method is just as reliable as one of the tools used in clinical research (RSTF) and more reliable than the clinical practice standard (CTF). \revision{Each VPT measurement using our smartphone staircase method takes $2$ to $3$ minutes to acquire, depending on how well the participant is able to perceive the vibrations. The smartphone-based exam thus takes longer to complete than the CTF or RSTF exams ($<20$ second completion time).} Next, we further evaluate the potential of the staircase method as a reliable tool when used without the support of an in-person experimenter.

\section{Study 2: Guided vs. Unguided}
\label{sec:guidance}
The purpose of this study was to better understand how users would interact with our tool when they do not have verbal clarifications and corrections from the experimenter (simulating how one may use the tool during an eventual at-home study). To that end, we measured VPTs in two usage conditions (unguided and guided) at two body locations (the hand and the foot) in 15 participants and compared VPTs for each condition. Below, we provide details on the smartphone measurement methods, user study protocol, and our statistical analysis followed by our results and discussion.

\subsection{Methods}
\subsubsection{Smartphone Measurement Methods}
We conducted a user study measuring the VPTs at two body positions (hand and foot) using the smartphone staircase method from the previous study comparing two usage conditions: (1) guided and (2) unguided. For this study, we made a few modifications to the vibration stimuli. Namely, we increased the length of the vibration (the Core Haptics variable ``\textit{hapticDuration}'') to $1.0$~s from the previous duration of 0.1~s. To accommodate for this change in our state machine, the interval between vibrations was randomly adjusted to a range of $4.5-7.5$~s instead of $3-6$~s. The period of time given to respond to each vibration was also increased to $3$~s from an original value of $2.5$~s.
These changes were based on input from our collaborating neurologists who suggested that the original vibration duration ($0.1$~s) was quite short, and that a value of $1.0$~s would be better for assessing vibration perception. 

\subsubsection{User Study Participants}
Fifteen adult participants with no known history of diabetes or other disorders linked to peripheral neuropathy completed this study. Self-reported demographics of the participants are as follows: 8 reported female and 7 male for their sex assigned at birth; 5 were between the ages of 18 to 23 (mean: 20.2, std: 0.84) and 10 between the ages of 40 to 58 (mean: 48.5, std: 4.38); all right-handed; 12 right-footed and 3 ambipedal; 2 were Black/of African descent, 2 were Asian, 7 were White, 5 were Hispanic or Latino, 1 was Hawaiian or other Pacific Islander, and 3 preferred not to answer (some participants listed multiple identities). The separate age groupings above are explicitly stated to provide granular information regarding the participant demographics since the range of ages was quite large. However, we do not make any claims regarding differences between the performance of the two age groups in this current study. This study was approved by the Stanford University Institutional Review Board under Protocol 22514, and written consent was provided by all participants. Prior to completing the study, participants completed a pre-survey that inquired about demographic information as well as hobbies or injuries that may impact touch sensitivity at the hands or feet.

\subsubsection{User Study Protocol}
Participants were provided an iPhone XS Max, with our app pre-loaded onto the phone. All participants completed the unguided condition first, followed by the guided condition. The order of hand or foot in each condition was randomized, and measurements at each location were repeated three times with one exception. The first location in the unguided condition was repeated four times so that users could familiarize themselves with the basic usage of the platform prior to beginning the experimental trials. 

In the unguided portion, participants were provided a printed set of instructions with the following step-by-step instructions: 

\begin{enumerate}
    \item Use the provided noise canceling headphones that play white noise with volume set at 75 percent. 
    \item Place the smartphone on the provided pillow and open the app. 
    \item Position your dominant index finger or big toe (experiment randomly assigned which location to do first) on the bottom center of the smartphone with the joint line located closest to the nail aligned with the bottom-center edge of the phone (Participants \revision{were} provided with Fig.~\ref{setup}).
    \item Click ``Next'' in the app to begin the trial with toes and fingers held in place. 
    \item Say ``yes'' each time a vibration is felt until the screen says that the trial is complete.
    \item Repeat the above process three times ensuring that the finger or toe remains in the designated position. 
\end{enumerate}

During this unguided phase, the experimenter took notes documenting how the user interacted with the phone and various points of confusion or error for the participants. The experimenter did not answer any questions to simulate how a participant might use the app during an in-home setting. We also recorded how the participant interacted with the phone (via video camera) so that it could be analyzed as necessary after the experiment. 

After completing the unguided condition, the experimenter answered any questions that the participants had. The experimenter also demonstrated how to place their finger and big toe on the phone. Then the participant completed 3 more trials at each location under the guidance of the experimenter who corrected any errors as they arose.

\begin{figure}[t]
\centering
\includegraphics[width=\linewidth]{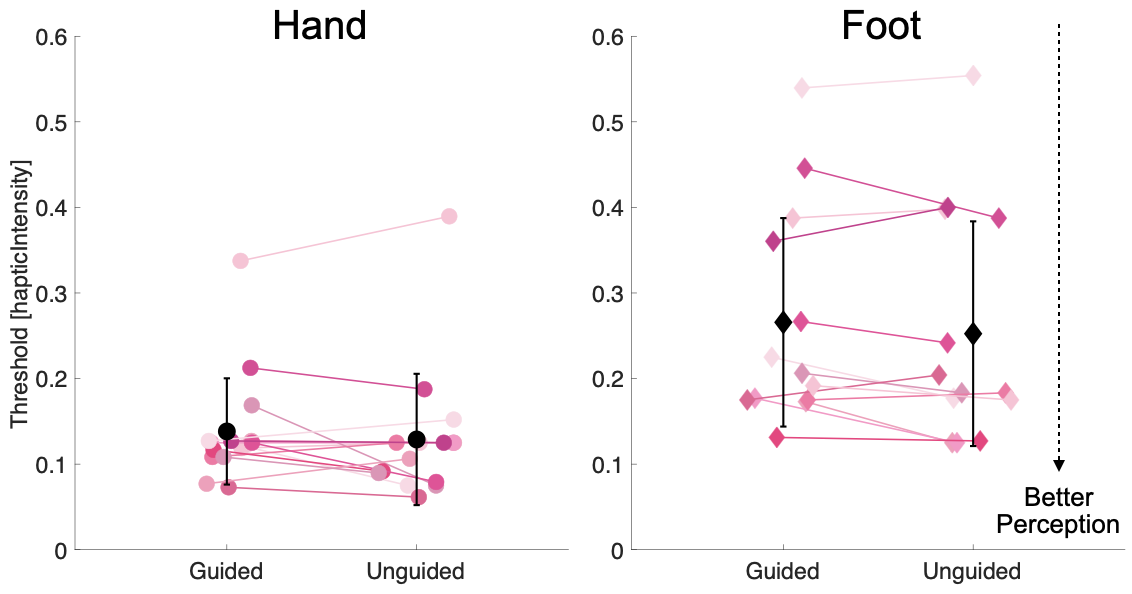}
\caption{Guided vs. unguided VPTs at the hand (left) and foot (right). Experimenter guidance did not significantly impact measured VPTs at the hand (p = 0.327) nor at the foot (p = 0.166). Individual subjects are shown in shades of pink (\revision{$n = 15$ for hand, $n = 13$ for foot}) and mean and standard deviation are shown in black.}
\label{bargraphs}
\end{figure}

\begin{figure*}[t]
\centering
\includegraphics[width=0.9\linewidth]{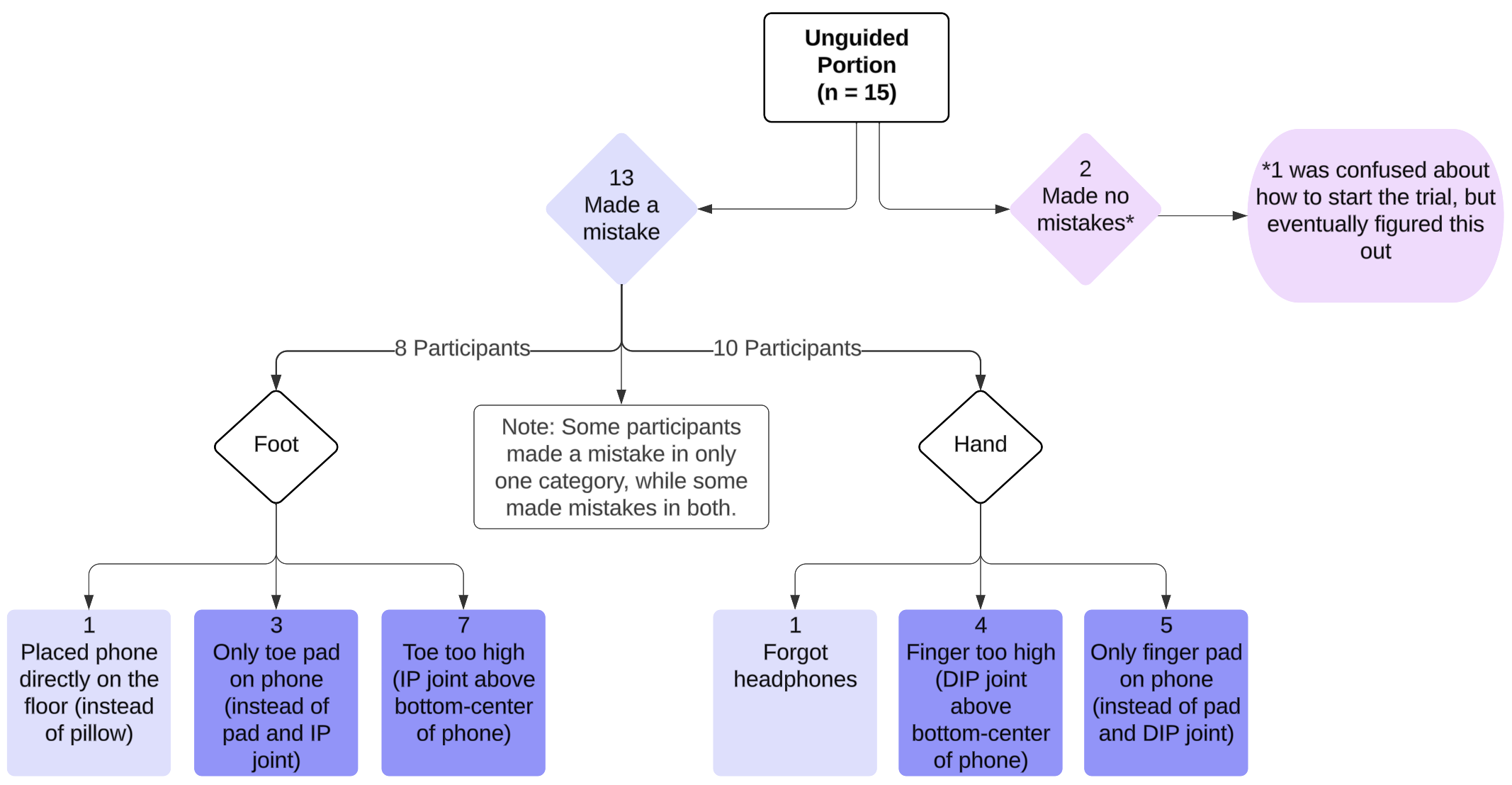}
\caption{Flowchart depicting the errors made by participants during the unguided condition. Participant errors predominately consisted of incorrect toe or finger placement, either placing only a portion of the finger or toe on the phone or placing the finger or toe too high on the phone\revision{ (noted in the darker-colored squares).} Some participants may have made multiple mistakes (for instance, having only the toe pad, while also placing it too high on the phone), while some participants may have made mistakes only for a specific trial.}
\label{observations}
\end{figure*}

\begin{figure}[h]
\centering
\includegraphics[width=\linewidth]{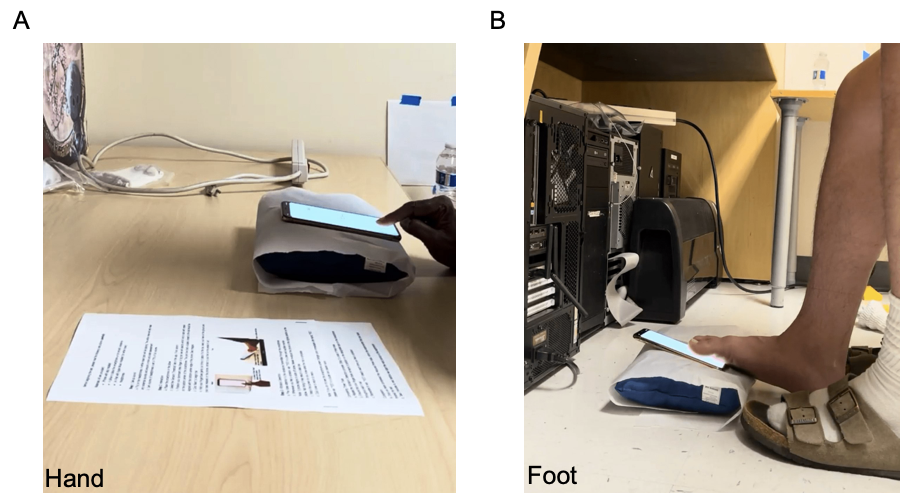}
\caption{The two most common errors made by participants in the unguided condition were (a) placing only the finger pad on the phone (instead of the finger pad and distal interphalangeal (DIP) joint) and (b) placing the toe too high on the phone (interphalangeal (IP) joint not aligned with bottom-center edge of phone). As a reminder, the instructed placements are found in Figure~\ref{setup}.}
\label{errors}
\end{figure}

\subsubsection{Statistical Methods}
Paired t-tests were used to compare the difference between the guided and unguided conditions at the hand and the foot. All analysis was conducted in R (Version 4.2.1) using R Studio (Version 2022.07.2). Notes taken by the experimenter during the trials as well as videos were evaluated to identify the most common errors in the unguided conditions.

Although data from a total of 15 participants was collected, two were removed from the quantitative analysis for the foot portion. One subject was removed from the foot portions because this subject couldn't feel any of the phone vibrations at the foot. The other subject was removed from the foot portion because, during the unguided portion of the foot trials, this subject placed the phone directly on the ground instead of on the pillow, which resulted in vibrations that were dampened to the point of not being able to be perceived by the participant. This participant felt the foot vibrations during the guided portion of the experiment when the experimenter was able to correct the user's mistake of placing the phone directly on the ground. \revision{As explained in Section~\ref{subsec:phone meas methods}, when a participant is unable to feel the maximum amplitude vibrations that the phone outputs (\textit{hapticIntensity} = $1$), their VPT for that trial is undefined and is thus recorded as \texttt{NaN}. As such, we cannot include these subjects in quantitative analyses (Fig.~\ref{bargraphs}), although we do include them in the qualitative analysis described in Fig.~\ref{observations}.}

\subsection{Statistical Results}

Figure~\ref{bargraphs} shows the individual subject data and mean and standard deviation of the measured VPTs for each body location when guided and unguided. Paired t-tests using the mean VPT of the three trials for each condition yielded a non-statistically significant difference between the guided and unguided conditions at the hand ($t(14) = -1.01$, $p = 0.327$, $d = -0.262$) or at the foot ($t(12) = -1.48$, $p = 0.166$, $d = -0.409$). The mean VPT for the unguided conditions for both the hand and foot were slightly lower (as indicated by the negative Cohen's d values), though not statistically significant.

\subsection {Observational Results}

Figure~\ref{observations} summarizes the errors that participants made during the study via a flowchart. Figure~\ref{errors} illustrates the most common errors that occurred when using the phone. The most common error for the foot was that the toe was placed too high on the phone (7 errors), and the most common error for the hand was that only the finger pad (and not the pad and distal interphalangeal (DIP) joint) were placed on the phone (5 errors). Similarly, the second most common error for the foot was that only the toe pad (and not the pad and interphalangeal (IP) joint) were placed on the phone (3 errors), and the second most common error for the hand was that the finger was placed too high on the phone (4 participants). Most of the mistakes in following the directions occurred in terms of finger or toe placement (9 hand, 10 foot), while only 2 errors total resulted from a procedural issue (such as forgetting to use headphones or placing the phone on the wrong surface).

\subsection{Discussion}

This lack of difference in VPT measurements between unguided and guided conditions at both the hand and the foot shows that guidance has minimal impact on VPT measured using the smartphone. This is promising, as it \revision{
suggests} that our smartphone app could eventually be distributed in unguided settings and yield reliable results. Although statistically insignificant, it is interesting that the measured VPTs from the unguided portion of the study resulted in slightly lower averages. This could be connected to the fact that many of the errors resulted from toe or finger placement being too high on the phone (which results in the limb having more contact with the phone). It is also known that accelerations vary throughout the phone~\cite{yoshida2023cognitive}, which may have resulted in different acceleration amplitudes that may have been more recognizable. To ensure that data is collected via consistent body placements even in unguided settings, we will modify the provided instructions to include an additional top view of the foot placement on the phone. We also plan to add an on-screen guide for foot placement (i.e.\ a green oval indicating where to place the toe). The foot location is of particular importance for diagnostic purposes such as tracking the progression of diabetic peripheral neuropathy~\cite{elsayed202312}. Hence any efforts we can make to improve data collected at the foot is of particular interest.

\section{Conclusion}
\label{sec:conclusion}
In this work, we conducted two separate user studies to investigate the reliability of smartphone-based vibration perception threshold measurements. From the first study, we found that our smartphone staircase method yielded repeatable measurements that did not significantly differ from one another and resulted in intraclass correlation coefficients indicating good or excellent reliability, on par with measurements from the Rydel-Seiffer tuning fork used in \revision{
neurological research} and better than the standard clinical tuning fork. We then conducted a second study using this smartphone staircase method to evaluate if the measured values would change if the participants were not given real-time guidance on how to use the tool, simulating how they might use the tool to conduct measurements on themselves at home. Excitingly, we discovered that the vibration perception threshold measurements collected when the participants received no guidance from an experimenter did not significantly differ from the measurements collected with assistance from an experimenter. We also classified common errors made by participants when conducting the measurements on themselves and presented modifications that can be made to the provided instructions and the smartphone interface to minimize these errors moving forward. The results from the work presented in this paper indicate that our smartphone staircase app could eventually be distributed for use in at-home, unguided settings for a variety of applications. Prior to realizing our long-term goal of distributing the staircase app to be used as a diagnostic tool, we will build upon this work by conducting more extensive testing in diverse groups of participants including both healthy and \revision{
patient} populations. \revision{
We will also} conduct tests in lab (enabling us to compare to neurology clinic gold standards such as electromyography-nerve conduction exams) and at home (allowing us to test the usability of the tool outside the lab or clinic). We are particularly excited by the potential of this tool to aid in measuring the progression and regression of sensory damage caused by various diseases such as diabetic peripheral neuropathy and chemotherapy-induced \revision{peripheral} neuropathy.  

\addtolength{\textheight}{-12cm}   




\section*{ACKNOWLEDGMENTS}
We thank the participants for their time and effort. We also thank our clinical collaborators Dr.\ Kenneth\ K.\ Leung and Dr.\ Srikanth\ Muppidi for their contributions to selection of the vibration stimuli.


\bibliographystyle{IEEEtran}
\bibliography{references.bib}
\end{document}